\documentclass[aps,showpacs,groupedaddress,10pt,pra]{revtex4-2}
\usepackage{amsmath}
\usepackage{amssymb}
\usepackage{hyperref}
\usepackage{graphicx}
\usepackage{epstopdf}
\usepackage{color}

\newcommand{\hodge}[1]{\,^*#1}
\begin{document}
\preprint{88}
\title{A study about black hole solutions with nonconstant transversal curvature and its conserved charges in Lovelock gravity.}

\author{Rodrigo Aros}
\email{raros@unab.cl}
\affiliation{Departamento de Ciencias Fisicas, Universidad Andres Bello, Av. Republica 252, Santiago,Chile}

\author{Milko Estrada }
\email{milko.estrada@gmail.com}
\affiliation{Facultad de Ingeniería y Empresa, Universidad Católica Silva Henríquez, Chile}

\date{\today}
\pacs{04.50.+h, 04.70.Bw}

\begin{abstract}
In this work, the analysis of some new static black hole solutions of Lovelock gravity with nonconstant curvature transverse section is presented. It will be shown that the finiteness of the charges and the action principle rely on the existence of constraints on the geometry of the transverse sections. Finally, in this context, some new sound solutions with nonconstant curvature transverse sections that deviate from the previously known geometries are discussed.

\end{abstract}

\maketitle

\section{Introduction}

There is no doubt that during the last decades the study of asymptotically AdS spaces, their corresponding conformal infinities, and conformal field theory have been mainstream topics in theoretical physics and mathematics. For some recent examples see \cite{blitz2023toward}. The main motivation of this has been the AdS/CFT correspondence \cite{Maldacena:1997re}, or holography in general \cite{Susskind:1994vu,Bousso:2002ju}.
For this work, however, the specific motivation is to establish the conditions satisfied by proper (asymptotically locally) AdS spaces  \cite{Hull:2021bry,Hull:2022xew} whose conformal infinities could be representatives of equivalent classes of conformal (differential) manifolds, given the evidence that non-trivial (conformal) ge ometries play a fundamental role in the study of conformal theories and their anomalies \cite{Liu:1998ty,Beccaria:2014xda,Beccaria:2017lcz,Mukherjee:2021alj}. It is worth to emphasize that this is not in conflict with the existence of static spaces, nor, in principle, with the Fefferman-Graham expansion \cite{Fefferman:2007rka} for asymptotic Einstein spaces. However, this indeed requires the extension of Birkhoff's theorem \cite{Ray:2015ava}.

To continue, it is worth mentioning that, as is known, not any theory of gravity has second-order EOM, and thus causality is not protected in general. This is usually puzzled out by merely ruling out the non-causal solutions. Furthermore, in the case of asymptotic AdS solutions are usually only considered those that are asymptotically Einstein spaces. However, this is not the final word, and one can indeed further extend the spectrum of \textit{proper} solutions by recalling the existence of other theories of gravity with second-order EOM, and thus to include solutions that do not converge into an Einstein space asymptotically. See \cite{Banados:1994ur,Crisostomo:2000bb,Aros:2000ij} and references therein for examples of these geometries. Finally, it must be noticed that the presence of some particular matter fields, such as scalar fields, could modify that asymptotic behavior and yet yield proper solutions. See for instance \cite{Martinez:1996gn} or \cite{Erices:2017izj}. For the vacuum solutions with constant curvature transverse sections, this scenario can be visualized by using Schwarzschild coordinates,
\begin{equation}\label{SchAdSCC}
  ds^2 = -f(r)^2 dt^2 + \frac{1}{f(r)^2} dr^2 + r^2(\bar{g}_{ij} dy^i dy^j),
\end{equation}
where $\bar{g}_{ij} dy^i dy^j$ stands for the line element of a transverse section of constant curvature $\Sigma_\gamma$ (with $\gamma=\pm 1,0$). In order to Eq.(\ref{SchAdSCC}) be an asymptotically locally AdS (\textbf{ALAdS}) space it must be satisfied that
\[
 \lim_{r\rightarrow \infty} f(r)^2 \sim \gamma + \frac{r^2}{l^2} - \frac{C_2}{r^a},
\]
with $a > 0$. For what follows it is worth to mention that Lovelock gravity \cite{Lovelock:1971yv} is the simplest case where the general statement above can be confirmed. In this case \cite{Crisostomo:2000bb,Aros:2000ij},
\begin{equation}\label{AsympFormSigma}
  \lim_{r\rightarrow \infty} f(r) \sim \gamma + \frac{r^2}{l^2} - \left(\frac{C_2}{r^{d-2k-1}}\right)^{1/k}.
\end{equation}

To finish it is worth recalling that extending the spectrum of proper solutions is not usually straightforward. This is because, on-shell, any proper solution must define a finite action principle and satisfy suitable boundary conditions that yield a well-defined variation principle. Furthermore, its associated conserved charges must be finite as well. The crux for any ALAdS space is that a regularization process must be introduced to attain these three conditions. Moreover, satisfying those conditions is mandatory to achieve a dual CFT interpretation within the AdS/CFT conjecture. Obviously, given the relevance of this problem, nowadays there are several different approaches to attain a regularized action principle. Among them, see for instance \cite{Skenderis:2002wp,Miskovic:2007mg,Miskovic:2014zja} and more recently \cite{Arenas-Henriquez:2017xnr,Arenas-Henriquez:2019rph,Anastasiou_2021,Anastasiou:2022wjq}.

In the next sections this work will analyze some new black hole solutions of Lovelock gravity with nonconstant curvature transverse section and the conditions under which the finiteness of the action principle and the conserved charges could be attained. Because of its general appliance, this work will follow the method of regularization described in \cite{Miskovic:2007mg,Mora:2006tq,Mora:2004kb}.

\subsection{Gravity}\label{Kgravities}

Among the many possible theories of gravity that are worth consideration, Lovelock gravities have a substantial role due to retaining in higher dimensions the essential features GR has. For instance, their equations of motion are second-order differential equations. The action principle on $d$-dimensional manifold $(\mathcal{M},g)$ is given by
\begin{equation}\label{Lovelock}
  \mathbf{L} = \sum_{p=0}^{[\frac{d-1}{2}]} \alpha_p L_p
\end{equation}
where $[X]$ stands for the integer part of $X$ and
\[
L_p = \frac{1}{2^p}\delta^{\mu_1 \ldots \mu_{2p}}_{\nu_1\ldots \nu_{2p}} R^{\nu_1 \nu_2}_{\hspace{2ex} \mu_1 \mu_2} \ldots  R^{\nu_{2p-1} \nu_{2p}}_{\hspace{2ex} \mu_{2p-1} \mu_{2p}} \sqrt{g},
\]
with $R^{\nu_1 \nu_2}_{\hspace{2ex} \mu_1 \mu_2}$  the Riemann tensor and $\{\alpha_p\}$ a set of arbitrary constants.

For shortness the Lovelock Lagrangian in Eq.(\ref{Lovelock}) can be written in terms of the Riemann two-form of curvature,  $R^{ab}=d\omega^{ab} + \omega^a_{\,\,c}\omega^{cb}$, and the vielbein $e^a$. See for instance \cite{Zanelli:2002qm}. In this formalism, the Lovelock action reads
\begin{equation}\label{initialaction}
{\mathbf{L}} = \sum_{p=0}^{n-1} \alpha_p \,\epsilon_{a_1\ldots a_{2n}} \left[(R)^{p}
\left(\frac{e}{l}\right)^{2n-2p}\right]^{a_1 \ldots a_{2n}},
\end{equation}
where
\[\left[(R)^{p} \left(\frac{e}{l}\right)^{2n-2p}\right]^{a_1 \ldots a_{2n}} = R^{a_1 a_2}\wedge \ldots
R^{a_{2p-1} a_{2p}} \wedge \frac{e}{l}^{a_{2p+1}} \wedge \ldots \wedge \frac{e}{l}^{a_{2n}}.
\]
The analysis of the corresponding equations of motion is depicted in appendix \ref{LovelockEOM}. Unfortunately, it is straightforward to show that the action principle above, either in even or odd dimensions, is ill-defined on an ALAdS space and thus it must be supplemented by a suitable boundary term $\Omega$ to attain a proper action principle. A suitable formalism in odd dimensions for the construction of $\Omega$ is described in appendix \ref{OddDReg}. See Refs.\cite{Miskovic:2007mg,Mora:2006tq,Mora:2004kb}. In even dimensions, this is much simpler and suffixes the addition of the corresponding Euler density \cite{Aros:1999kt}.

To continue, the form of the Noether current, associated with the invariance under diffeomorphisms $x \rightarrow x + \xi$, is given by
\begin{equation}\label{GravitationalCharge}
   \hodge{{\bf J}_\xi} = -d\left(I_\xi w^{ab} \tau_{ab} + I_\xi \Omega\right),
\end{equation}
where
\[
\tau_{ab} =  \frac{\partial \mathbf{L}}{\partial R^{ab}}.
\]

\subsection{Horizon, Killing vectors and Boundary Conditions}\label{Bconditions}

Let $\mathcal{M}$ be a static black hole geometry of topology ${\mathbb{R}} \times \Sigma$. Being $\mathcal{M}$ a black hole geometry it must have an internal boundary to accommodate the presence of a horizon, \textit{i.e.}, $\partial \mathcal{M} = \partial \mathcal{M}_{\infty} \oplus \partial \mathcal{M}_{H}$. For simplicity, it will be assumed that $\partial \Sigma = \partial\Sigma_\infty \oplus \partial\Sigma_H$, which denotes the spatial infinity and the event horizon respectively. The boundary of ${\mathcal{M}}$ is therefore given by
\[
\partial{\mathcal{M}}=\partial \Sigma \times {\mathbb{R}}=\partial \Sigma_H \times
{\mathbb{R}}\cup\partial \Sigma_\infty \times {\mathbb{R}}.
\]

As mentioned above, $\hodge{{\bf J}_\xi}$ can be constructed for any $\xi$. However, to define a physical conserved charge some conditions must be satisfied. First, it is necessary that $\mathcal{M}$, has, at least asymptotically, a timelike symmetry. Second, $\xi$ must generate an isometry, namely a Killing vector. Furthermore, $\xi$ must be compatible with preserving the boundary conditions. With all this in mind, following \cite{Jacobson:1993xs,Wald:1993nt,Wald:1984rg}, $\xi$ will be considered the null generator of the event horizon \cite{hawking}. This implies, in first-order formalism, that
\begin{equation}\label{SurfaceGrav}
    \left.I_{\xi}\omega^a_{\hspace{1ex} b} \xi^b \right|_{\partial \Sigma_H}=  \left. \kappa \xi^a \right|_{\partial \Sigma_H}
\end{equation}
where $\kappa = 4\pi T$ is the surface gravity with $T$ the temperature of the horizon. Notice that fixing $\omega^{ab}$ at the horizon fixes $T$. Furthermore, it must be stressed that fixing $\omega^{ab}$ is a suitable boundary condition at the horizon and equivalent to fixing the second fundamental form or the extrinsic curvature on the horizon. See \cite{choquet1982analysis}.

Now, considering the asymptotic region, any asymptotically local AdS space satisfies
\begin{equation}\label{AsympT}
\lim_{x \rightarrow \partial \mathcal{M}_{\infty}} R^{\mu\nu}_{\hspace{2ex}\alpha\beta} \rightarrow -\frac{1}{l^2} \delta^{\mu\nu}_{\alpha\beta},
\end{equation}
where $l$ represents an effective AdS radius.

To continue with the discussion, let be a Schwarzschild ansatz with coordinate system $x^{\mu} = (t,r,y^i)$,
\begin{equation}\label{SchAdS}
  ds^2 = -f(r) dt^2 + \frac{1}{f(r)} dr^2 + r^2(\hat{g}_{ij} dy^i dy^j),
\end{equation}
where $\hat{g}_{ij} dy^i dy^j$ is the line element of an arbitrary transverse section $\Sigma$. It is straightforward to observe that $f(r)\geq 0$ determines a well-defined region, being $\xi = \partial_t$ the null generator of the horizon due to $\xi \cdot \xi = -f(r)$. In this way, $\partial \Sigma_{H}$, is defined by a \textit{radius} $r=r_+$ subjected to $f(r_+)=0$. If $f(r)$ has more than one root, it will be assumed that $r=r_+$ is the largest one.

The asymptotic region is defined by $r\rightarrow \infty$ and would be considered an asymptotically locally AdS. The condition (\ref{AsympT}) determines, to leading orders, that
\begin{equation}\label{GeneralAsymptotiaSO}
\lim_{r \rightarrow \infty}f(r) \sim \Gamma + \frac{r^2}{l^2} + O(r^{-a}),
\end{equation}
where $a>0$ and $\Gamma$ a universal constant to be determined \cite{Fefferman:2007rka}.

\subsection{Beyond the constant curvature}
To study spaces with nonconstant curvature transverse sections $\Sigma$, it is convenient to define the set of constants $c(q)$ \cite{Ray:2015ava}
\begin{eqnarray}
c(q) &=& \int_{\Sigma} \frac{(d-2q-2)!}{2^q}\delta^{i_1 \ldots i_{2q}}_{i_1\ldots i_{2q}} \hat{R}^{i_1 i_2}_{\hspace{2ex} j_1 j_2} \ldots  \hat{R}^{i_{2q-1} i_{2q}}_{\hspace{2ex} j_{2q-1} j_{2q}} \sqrt{\hat{g}} d^{d-2}y \label{c(q)} \\
   &=&  \int_{\Sigma} \varepsilon_{a_1\ldots a_{d-2}} \hat{R}^{a_1 a_2} \ldots  \hat{R}^{a_{2q-1} a_{2q}} \hat{e}^{a_{2q+1}} \ldots \hat{e}^{a_{d-2}}
\end{eqnarray}
where $q=0,\ldots [(d-2)/2]$ and $ \hat{R}^{i_1 i_2}_{\hspace{2ex} j_1 j_2} $ stands for the Riemann tensor of $\Sigma$.
Firstly, it can be noticed that
\[
c(0) = (d-2)! Vol(\Sigma) \textrm{ with }  Vol(\Sigma) \textrm{ finite.}
\]

\section{Gauss Bonnet Gravity}\label{GaussBonnetSection}

As an introduction to the analysis, this section will address the solutions of Gauss-Bonnet gravity with nonconstant transversal curvature. This was original presented in \cite{Hull:2021bry}.  It is worth to mention that in the construction of these solutions was introduced a set of $b(n)$ coefficients which differ from the $c(q)$ in eq.(\ref{c(q)}) by a normalization factor. As new results, in this section, it will be computed the Noether charge for $d=5$ and the necessary conditions to have a finite Noether charge and action principle for $d>5$.

Einstein-Gauss-Bonnet (EGB) gravity is defined by arbitrary $\alpha_p$  for $p=0,1,2$ and $\alpha_p =0$  $\forall p > 2$. {For the constant transversal curvature case,} as {was} discussed in \cite{Banados:1994ur,Crisostomo:2000bb}, this theory has ALAdS solutions provided certain conditions are satisfied. The solution, for $d \geq 5$, is given by
\begin{equation}\label{TheSolution}
    f(r) = \frac{b(1)}{b(0)} + \frac{\alpha_1}{2 \alpha_2}r^2 - \frac{\sqrt{4 \alpha_2^2 (b(1)^2 - b(2)b(0)) + b(0)^2(\alpha_1^2-4\alpha_0\alpha_2)r^4 + {\displaystyle \frac{8 m \alpha_2 b(0)  }{r^{d-5}}}}}{2 b(0) \alpha_2}.
\end{equation}
which coincides with Ref.\cite{Hull:2021bry} for $b(0)=\alpha_1=1$ and
\[
b(1) = \frac{(d-4)!}{2(d-2)!}\int \hat{R}\sqrt{\hat{g}} d^{d-2}y \textrm{  and  } b(2) = \frac{(d-6)!}{4(d-2)!}\int \delta^{i_1 \ldots i_{4}}_{i_1\ldots i_{4}} \hat{R}^{i_1 i_2}_{\hspace{2ex} j_1 j_2} \hat{R}^{i_{3} i_{4}}_{\hspace{2ex} j_{3} j_{4}} \sqrt{\hat{g}} d^{d-2}y,
\]
respectively.

The asymptotic form Eq.(\ref{TheSolution}) for $d\ge 5$ is given by
\begin{eqnarray}
   \lim_{r\rightarrow \infty}  f(r)  &=& \frac{b(1)}{b(0)} + \frac{\alpha_1}{\alpha_2}\left(1 - \sqrt{1-4\frac{\alpha_0\alpha_2}{\alpha_1^2}}\right) r^2 -\left(\left(\frac{b(1)}{b(0)}\right)^2 \frac{\alpha_2}{\alpha_1}\frac{ (b(2)b(0) - b(1)^2)}{\sqrt{1-4\frac{\alpha_0\alpha_2}{\alpha_1^2}}}\right) \frac{1}{r^2} \nonumber \\
   &-& \frac{2m}{b(0) \alpha_1 \sqrt{1 - 4\frac{\alpha_0 \alpha_2}{\alpha_1^2} }} \frac{1}{r^{d-3}} + \ldots\label{TheSolutionAsymp}
\end{eqnarray}
Here, if this solution is to describe an ALAdS space, it becomes mandatory that $\alpha_1^2 > 4\alpha_0\alpha_2$. This yields an effective AdS radius given by
\[
\frac{1}{l^2_{\text{eff}}} =  \frac{1}{l^2} = \frac{\alpha_1}{\alpha_2}\left(1 - \sqrt{1-4\frac{\alpha_0\alpha_2}{\alpha_1^2}}\right).
\]

One can also notice that $b(1)/b(0)$ had replaced the value of the constant curvature of the transverse section in Eq.(\ref{AsympFormSigma}) \cite{Crisostomo:2000bb}.

Continuing, the general behavior mentioned of Eq.(\ref{GeneralAsymptotiaSO}) is satisfied by Eq.(\ref{TheSolutionAsymp}), as expected. Next, beyond the first two terms, one can also notice the presence of a term of $O(r^{-2})$, and one of $O(r^{3-d})$, the later the one expected for the Schwarzschild solution in $d$ dimensions. This hints that the term $O(r^{-2})$ must be removed for $d>5$. To confirm this, the charges and action principle will be evaluated on this solution.

\subsection{Five dimensions}
As mentioned above, in $d=5$ the asymptotic form, see Eq.(\ref{TheSolutionAsymp}), does match the asymptotia of GR, but also $b(2)=0$. The conserved charge associated with $\xi = \partial_t$ is finite and given by
\begin{equation}\label{Mass}
  Q(\partial_t) = E = m  + \frac{b(1)^2}{8 b(0)^2}\left(2 \alpha_2-l_{\text{eff}}^2 \alpha_1\right).
\end{equation}

Here, one can notice that $E$ corresponds to the mass $m$ plus the \textit{would-be vacuum energy} of the solution. This is the generalization of the result obtained in \cite{Mora:2004rx} in the case of a constant curvature transversal section. The action principle is also finite and given
by
\begin{equation}\label{ActionFinite}
  I_{reg} = - \frac{b(1)^2}{4b(0)}l^2_{\text{eff}}(2l^2_{\text{eff}}\alpha_0+1).
\end{equation}
This shows that for $d=5$ the finiteness of both action principle and charges, for the EGB solution with nonconstant transversal curvature, can be attained for any $b(1) \in \mathbb{R}$ as expected due to the asymptotic form (\ref{TheSolutionAsymp}) does indeed match the asymptotia of GR.

\subsection{More than five dimensions}
In $d>5$ $b(2)\neq 0$ and the Noether charge $Q(\partial_t)$, computed following Eq.\ref{NoetherOddFormal}, diverges as
\begin{equation}
  \lim_{r \rightarrow \infty} Q(\partial_t) \approx (b(2)b(0) - b(1)^2) r^{d-5} + \textrm{finite} + \ldots.
\end{equation}
By computing the action principle it can be checked that this also diverges by a term proportional to $b(2)b(0) - b(1)^2$. Therefore, a proper action principle can be attained provided
\begin{equation}\label{HardCurvatureCondition}
 b(2) = \frac{b(1)^2}{b(0)}.
\end{equation}
This condition is trivially yielded by any constant curvature transverse section.

It must be noticed that Eq. (\ref{HardCurvatureCondition}) represents a nontrivial constraint for any non-constant curvature manifold. This also confirms that the term of O$(r^{-2})$ for $d>5$ represents an obstruction to be removed to have sound and meaningful solutions.

\section{Higher order Lovelock gravities}
To continue the analysis of the EOM, and to address the higher-order Lovelock gravities, it is necessary to unveil a different structure of the EOM. Firstly, it is necessary to manifest the presence of the $c(q)$ mentioned above. Second, one can notice that the EOM, for $\alpha_p = 0$ for $p>q$, can be expressed in the \textit{pseudo-polynomial} fashion
\begin{equation}\label{LEOM}
  (\mathcal{E}_{d}^q)^{\alpha}_{\beta} = \alpha_q \delta_{\beta \mu_1 \ldots \mu_{2q}}^{\alpha \nu_1 \ldots \nu_{2q}} (R^{\mu_1 \mu_2}_{\nu_1 \nu_2}+\beta_1 \delta^{\mu_1 \mu_2}_{\nu_1 \nu_2} )\ldots (R^{\mu_{2q-1} \mu_{2q}}_{\nu_{2q-1} \nu_{2q}}+\beta_{q} \delta^{\mu_{2q-1} \mu_{2q}}_{\nu_{2q-1} \nu_{2q}}) =0,
\end{equation}
where $q \in [1\ldots [(d-1)/2]]$ depends on the particular Lovelock gravity considered. It must be stressed that $\beta_i$'s coefficients cannot be obtained from the $\alpha_p$'s in general for $d \geq 9$ \footnote{The $\beta_i$ coefficients could be obtained by computing the zeros of the indexical equation, on a variable $\zeta$, got by replacing Riemann tensor in Eq.(\ref{Lovelock}) by $\zeta \delta^{\alpha\beta}_{\mu\nu}$. Unfortunately, this cannot be done in general for $d \geq 9$ due to the roots of a general polynomial of order higher than 4 cannot be obtained}. Fortunately, see below, the cases of interest can be analyzed without much ado until 8 dimensions.

Before proceeding it is worth to stress that among the trivial/ground state solutions of Eqs.(\ref{LEOM}), namely $R^{\mu \nu}_{\hspace{2ex}\alpha \beta}=-\beta_{i} \delta^{\mu\nu}_{\alpha\beta}$, only the subset $\beta_i > 0$ is of interest in this work, as those are locally AdS spaces. $\beta_{i} < 0$ and $\beta_i=0$, represent the locally de Sitter and flat solutions. Unfortunately, their analyses cannot be extrapolated from the one presented here. On top of that, there are also potentially (trivial) nonphysical solutions to Eq.(\ref{LEOM}), since starting from $\forall \alpha_p \in \mathbb{R}$, in general, some of the $\beta_i$'s might be complex numbers with non-vanishing imaginary parts.

\section{Higher order AdS equations}

In order to simplify the analysis it will set $\beta_i = l^{-2}>0$ to have the familiar form
\begin{equation}\label{AsymptAdSBehavior}
\lim_{x \rightarrow \partial M_{\infty}} R^{\mu\nu}_{\hspace{2ex}\alpha\beta} \rightarrow -\frac{1}{l^2}. \delta^{\mu\nu}_{\alpha\beta}.
\end{equation}
In general, some of the $\beta_i$ can be repeated and thus EOM can present degeneration around an AdS ground state. This has been studied previously in a different way in \cite{Crisostomo:2000bb} for spaces with constant curvature transversal sections. To address this situation is useful to reshape the EOM into the form
\begin{equation}\label{EOMAdSForm}
(\mathcal{E}_{d}^{(q,k)})^{\alpha}_{\beta} =  \alpha_q   \left(R + \frac{\delta}{l^2}\right)^{\mu_1 \ldots \mu_{2k}}_{\nu_1 \ldots \nu_{2k}} \left(\sum_{j=0}^{q-k}  K_j \left((2)^{2(q-k-j)} \delta_{\beta \mu_1 \ldots \mu_{2k+2j}}^{\alpha \nu_1 \ldots \nu_{2k+2j}}R^{\mu_{2k+1} \mu_{2k+2}}_{\nu_{2k+2} \nu_{2k+2}} \ldots R^{\mu_{2k+2j-1} \mu_{2k+2j}}_{\nu_{2k+2j-1} \nu_{2k+2j}} \right)\right) = 0,
\end{equation}
where $\{K_j\}$ is a set of dimensional constants and
\[
\left(R + \frac{\delta}{l^2}\right)^{\mu_1 \ldots \mu_{2k}}_{\nu_1 \ldots \nu_{2k}}  = \left( R^{\mu_{1} \mu_{2}}_{\nu_{1} \nu_{2}} + \frac{1}{l^2} \delta^{\mu_{1} \mu_{2}}_{\nu_{1} \nu_{2}}\right)\ldots \left( R^{\mu_{2k-1} \mu_{2k}}_{\nu_{2k-1} \nu_{2k}} + \frac{1}{l^2} \delta^{\mu_{2k-1} \mu_{2k}}_{\nu_{2k-1} \nu_{2k}}\right).
\]
It can be noticed that the degeneration and the AdS asymptotia of the solutions are both manifest in Eq.(\ref{EOMAdSForm}). It is worth mentioning that, unlike for the set $\{\beta\}$,  see \cite{Aros:2019vpu}, there is a well-defined relation between $\alpha_p$ and $K_i$ given by
\begin{equation}\label{AlphaCoeff}
  \alpha_p = \alpha_q \frac{1}{d-2p}\sum^{[d/2]-k}_{i=0} \binom{k}{p-i} K_i.
\end{equation}
for $p \leq q$ and $\alpha_p=0$ for $p>q$.

\subsection*{The static ansatz}
The use of the static ansatz displayed above, see Eq.(\ref{SchAdS}), simplifies the equations significatively. For instance, this implies that $L_p$, Eq.(\ref{Lovelock}), can be written as
\begin{eqnarray}
  \int  L_p &=& \int dt\wedge dr \left( \frac{d^2}{dr^2} \sum_{q=0}^{p} \binom{p}{q}(-f)^{p-q} c(q) \left(\frac{r}{l}\right)^{d-2p} \right)\nonumber\\
            &=& \int dt  \left.\left( \frac{d}{dr} \sum_{q=0}^{p} \binom{p}{q} (-f)^{p-q}c(q)\left(\frac{r}{l}\right)^{d-2p}\right) \right|_{r=r_+}^{r_{\text{max}}}\label{LLonAnsatz},
\end{eqnarray}
with the $c(q)$ are given by Eq.(\ref{c(q)}). Since it is satisfied by definition that $f(r_+)=0$ then
\[
\left.\left( \frac{d}{dr} \sum_{q=0}^{p} \binom{p}{q} (-f)^{p-q}c(q)\left(\frac{r}{l}\right)^{d-2p}\right) \right|_{r=r_+} = c(p-1)  p \frac{df(r_+)}{dr} \left(\frac{r_+}{l}\right)^{d-2p} + c(p) \frac{d-2p}{l} \left(\frac{r_+}{l}\right)^{d-2p-1}
\]
This result gives rise to the operational definition of
\begin{eqnarray*}
    P(r_+) &=&  \beta \sum_{p=0}^{[(d-2)/2]} c(p-1)  p \frac{df(r_+)}{dr} \left(\frac{r_+}{l}\right)^{d-2p} + c(p) \frac{d-2p}{l} \left(\frac{r_+}{l}\right)^{d-2p-1} \\
    &=&   \sum_{p=0}^{[(d-2)/2]} 4\pi c(p-1)  p  \left(\frac{r_+}{l}\right)^{d-2p} + \beta c(p)\frac{d-2p}{l} \left(\frac{r_+}{l}\right)^{d-2p-1}
\end{eqnarray*}
where $\beta = 4\pi \left( \frac{df(r_+)}{dr}\right)^{-1}$ is the Euclidean period. $c(-1)=0$ by definition.

The equations of motion (\ref{EOMAdSForm}) can be also written in a relatively simple form. For instance,
\begin{equation}\label{EOMAdSForm2}
   (\mathcal{E}_{d}^{(q,k)})^{0}_{0} \sim  \alpha_q\sum_{i=0}^{q-k}  K_i  \left( \sum_{s=0}^{k}\sum_{t=0}^{i} c(s+t) \binom{i}{t} \binom{k}{s} \frac{d}{dr}\left(\left(\frac{r^2}{l^2}-f(r)\right)^{k-s} (-f(r))^{i-t} \left(\frac{r}{l}\right)^{d-2k-2i-1}\right)\right).
\end{equation}
The rest of the components present a similar, and compatible, structure. Direct integration of these EOM yields
\begin{equation}\label{EOMAdSSolved}
     \alpha_q\sum_{i=0}^{q-k}  K_i  \left( \sum_{s=0}^{k}\sum_{t=0}^{i} c(s+t) \binom{i}{t} \binom{k}{s} \left(\left(\frac{r^2}{l^2}-f(r)\right)^{k-s} (-f(r))^{i-t} \left(\frac{r}{l}\right)^{d-2k-2i-1}\right)\right) = C,
\end{equation}
with $C$ an arbitrary integration constant.

A noteworthy feature of these equations (\ref{EOMAdSSolved}) is that their higher power can only increase each time a new odd dimension is reached \footnote{Roughly speaking, since the higher power of the Riemann curvatures can only increase each time a new odd dimension is reached, then the higher power}. Therefore, the highest power on $f(r)$, see Eq.(\ref{EOMAdSSolved}), of this equation in even dimensions must coincide with one of the odd dimensions below.

Naively it seems that obtaining $f(r)$ would only require solving Eq.(\ref{EOMAdSSolved}). Unfortunately, $f(r)$ can only be obtained \textbf{algebraically} if the order of Eq.(\ref{EOMAdSSolved}) is lower than 5, \textit{i.e.}, if and only if $q \leq 4$. This restricts the general case to dimensions lower than 9. It must be emphasized that this does not mean the lack of solutions for $d \geq 9$, but that solution can only be obtained for particular sets of coefficients. Because of this, in what follows, only the five and seven-dimensional cases, and some of their extensions, will be discussed in detail. The even dimensional $d\leq 8$ cases will be omitted because they are direct from the odd-dimensional cases for $d \leq 7$. As mentioned above, this is due to their corresponding equations, see Eq.(\ref{EOMAdSSolved}), which contain the same powers in $f(r)$ and only differ in the power of $r$.

\section{Five dimensions}

Before continuing it could be useful to recall that in $d=5$ there are three Lovelock theories to consider. Their corresponding equations of motion are given by
\begin{eqnarray}
     K_0 \alpha_1 \delta^{\alpha \nu_1 \nu_{2}}_{\beta \mu_1 \mu_{2}}  \left( R^{\mu_{1} \mu_{2}}_{\nu_{1} \nu_{2}} + \frac{1}{l^2} \delta^{\mu_{1} \mu_{2}}_{\nu_{1} \nu_{2}}\right)\ &=& 0 \label{EH5d}\\
   \alpha_2  \delta^{\alpha \nu_1 \ldots \nu_{4}}_{\beta \mu_1 \ldots \mu_{4}}  \left( R^{\mu_{1} \mu_{2}}_{\nu_{1} \nu_{2}} + \frac{1}{l^2} \delta^{\mu_{1} \mu_{2}}_{\nu_{1} \nu_{2}}\right) \left( K_1 R^{\mu_{3} \mu_{4}}_{\nu_{3} \nu_{3}} + K_0 \delta^{\mu_{3} \mu_{4}}_{\nu_{3} \nu_{4}}\right)\  &=& 0 \label{LL5Dk1} \\
     K_0 \alpha_2 \delta^{\alpha \nu_1 \ldots \nu_{4}}_{\beta \mu_1 \ldots \mu_{4}}  \left( R^{\mu_{1} \mu_{2}}_{\nu_{1} \nu_{2}} + \frac{1}{l^2} \delta^{\mu_{1} \mu_{2}}_{\nu_{1} \nu_{2}}\right) \left( R^{\mu_{3} \mu_{4}}_{\nu_{3} \nu_{4}} + \frac{1}{l^2} \delta^{\mu_{3} \mu_{4}}_{\nu_{3} \nu_{4}}\right) &=& 0 \label{CS5d}
\end{eqnarray}
Here Eq.(\ref{EH5d} are the EOM of $5d$ general relativity. The Eqs.(\ref{LL5Dk1}) correspond to the case EH action plus a general Gauss-Bonnet term, already discussed in section \ref{GaussBonnetSection}. Finally, it can be also recognized that the Eqs.(\ref{CS5d}) are the EOM of Chern-Simons gravity \cite{Hassaine:2004pp}.

As mentioned before, the intention is to analyze the solutions along the branches whose asymptotic behavior is given by $R^{\mu_{1} \mu_{2}}_{\nu_{1} \nu_{2}} \rightarrow - \frac{1}{l^2} \delta^{\mu_{1} \mu_{2}}_{\nu_{1} \nu_{2}}$ for nonconstant curvature transverse sections. As discussed previously, in $d=5$ Eq.(\ref{TheSolution}) is the solution to Eq.(\ref{LL5Dk1}) with the desired asymptotia,
\[
\lim_{r\rightarrow \infty} f(r) \sim \frac{c(1)}{c(0)} + \frac{r^2}{l^2} - \frac{C}{r^2},
\]
but has no restrictions on the values of $c(q)$ in $d=5$ due to $c(q)=0$ for $q>1$. {As was computed above, the energy for this case is given by equation \eqref{Mass}}.

\subsection*{Chern Simons}

Eq.(\ref{CS5d}) represents $d=5$ Chern Simons gravity, see \cite{Banados:1994ur}. This an independent case not covered by the discussion in section \ref{GaussBonnetSection}. To our knowledge this has not been discussed so far in the literature for nonconstant transversal curvature. Ths solution can be obtained from the relation (\ref{EOMAdSSolved}) which in this case is given by
\begin{equation}\label{CSsolutionAlg}
  \left(\frac{r^2}{l^2} - f(r) \right) c(1) + \frac{1}{2}\left(\frac{r^2}{l^2} - f(r) \right)^2 c(0) = M,
\end{equation}
where $M$ is an integration constant. The solution is given by
\begin{equation}\label{CSsolution}
  f(r) = \frac{c(1)}{c(0)} + \frac{r^2}{l^2} - \sqrt{\left(\frac{c(1)}{c(0)}\right)^2-\frac{2M}{c(0)}}.
\end{equation}

The process of regularization to compute the energy can be carried out following the standard procedure depicted in the appendix \ref{OddDReg}. The result is given by
\begin{equation}
Q\left(\partial_t \right)= K_0 M l^2 - K_0 \frac{c(1)^2}{2c(0)} l^2 = M l^2 + E_{\text{vacuum}},
\end{equation}
where the \textit{mass}, $M l^2$, and the energy of the vacuum $E_{\text{vacuum}}$ has been split. The presence of vacuum energy is a known fact of AdS gravity. See Ref. \cite{Mora:2004rx}.

It must be stressed that, as previously mentioned for GR and GB-GR, neither the regulation process nor the finiteness of the energy, impose any restriction on the values of $c(1)$ and $c(0)$.

\section{Seven and higher dimensions}

In seven dimensions there are a significantly larger number of cases to consider. These are
\begin{eqnarray}
     K_0 \alpha_1 \delta^{\alpha \nu_1 \nu_{2}}_{\beta \mu_1 \mu_{2}}  \left( R^{\mu_{1} \mu_{2}}_{\nu_{1} \nu_{2}} + \frac{1}{l^2} \delta^{\mu_{1} \mu_{2}}_{\nu_{1} \nu_{2}}\right)\ &=& 0 \label{EH7d}\\
   \alpha_2  \delta^{\alpha \nu_1 \ldots \nu_{4}}_{\beta \mu_1 \ldots \mu_{4}}  \left( R^{\mu_{1} \mu_{2}}_{\nu_{1} \nu_{2}} + \frac{1}{l^2} \delta^{\mu_{1} \mu_{2}}_{\nu_{1} \nu_{2}}\right) \left( K_1 R^{\mu_{3} \mu_{4}}_{\nu_{3} \nu_{3}} + K_0 \delta^{\mu_{3} \mu_{4}}_{\nu_{3} \nu_{4}}\right)\  &=& 0 \label{LL7Dk1} \\
     K_0 \alpha_2 \delta^{\alpha \nu_1 \ldots \nu_{4}}_{\beta \mu_1 \ldots \mu_{4}}  \left( R^{\mu_{1} \mu_{2}}_{\nu_{1} \nu_{2}} + \frac{1}{l^2} \delta^{\mu_{1} \mu_{2}}_{\nu_{1} \nu_{2}}\right) \left( R^{\mu_{3} \mu_{4}}_{\nu_{3} \nu_{4}} + \frac{1}{l^2} \delta^{\mu_{3} \mu_{4}}_{\nu_{3} \nu_{4}}\right) &=& 0 \label{LLk27d}\\
     \alpha_3 \delta^{\alpha \nu_1 \ldots \nu_{6}}_{\beta \mu_1 \ldots \mu_{6}}  \left( R^{\mu_{1} \mu_{2}}_{\nu_{1} \nu_{2}} + \frac{1}{l^2} \delta^{\mu_{1} \mu_{2}}_{\nu_{1} \nu_{2}}\right) \left( K_2 R^{\mu_{3} \mu_{4}}_{\nu_{3} \nu_{4}}R^{\mu_{5} \mu_{6}}_{\nu_{5} \nu_{6}}  + K_1 R^{\mu_{3} \mu_{4}}_{\nu_{3} \nu_{4}} \delta^{\mu_{5} \mu_{6}}_{\nu_{5} \nu_{6}} +  K_0 \delta^{\mu_{3} \mu_{4}}_{\nu_{3} \nu_{4}} \delta^{\mu_{5} \mu_{6}}_{\nu_{5} \nu_{6}}\right) &=& 0 \label{LLk17d2}\\
     \alpha_3 \delta^{\alpha \nu_1 \ldots \nu_{6}}_{\beta \mu_1 \ldots \mu_{6}}  \left( R^{\mu_{1} \mu_{2}}_{\nu_{1} \nu_{2}} + \frac{1}{l^2} \delta^{\mu_{1} \mu_{2}}_{\nu_{1} \nu_{2}}\right) \left( R^{\mu_{3} \mu_{4}}_{\nu_{3} \nu_{4}} + \frac{1}{l^2} \delta^{\mu_{3} \mu_{4}}_{\nu_{3} \nu_{4}}\right) \left( K_1 R^{\mu_{5} \mu_{6}}_{\nu_{5} \nu_{6}} + K_0 \delta^{\mu_{5} \mu_{6}}_{\nu_{5} \nu_{6}}\right) &=& 0 \label{LLk27d2}\\
      K_0 \alpha_3 \delta^{\alpha \nu_1 \ldots \nu_{6}}_{\beta \mu_1 \ldots \mu_{6}}  \left( R^{\mu_{1} \mu_{2}}_{\nu_{1} \nu_{2}} + \frac{1}{l^2} \delta^{\mu_{1} \mu_{2}}_{\nu_{1} \nu_{2}}\right) \left( R^{\mu_{3} \mu_{4}}_{\nu_{3} \nu_{4}} + \frac{1}{l^2} \delta^{\mu_{3} \mu_{4}}_{\nu_{3} \nu_{4}}\right) \left( R^{\mu_{5} \mu_{6}}_{\nu_{5} \nu_{6}} + \frac{1}{l^2} \delta^{\mu_{5} \mu_{6}}_{\nu_{5} \nu_{6}}\right) &=& 0 \label{CS7d}
\end{eqnarray}

Here Eq.\eqref{EH7d} corresponds to General Relativity in $d=7$, as Eq(\ref{EH5d}) does in $d=5$. Equations \eqref{LL7Dk1}  and \eqref{LLk27d} correspond to the EGB gravity with different couplings, with Eqs.(\ref{LL7Dk1}) representing the GB gravity already discussed in section \ref{GaussBonnetSection}. Finally, Equations \eqref{LLk17d2}, \eqref{LLk27d2}, and \eqref{CS7d} correspond to cubic gravities with different coupling constants. In particular Eqs.(\ref{CS7d}) represents  Chern Simons gravity in seven dimensions. These were initially discussed for constant transversal curvature in \cite{Banados:1994ur}.

Before returning to the discussion of the solutions with nonconstant curvature transverse sections, it is worth recalling that for solutions with constant curvature transverse sections is known that
\begin{itemize}
  \item the three Eqs.(\ref{EH7d},\ref{LL7Dk1}) and (\ref{LLk17d2}) share a branch satisfying Eq.(\ref{AsymptAdSBehavior}) and whose solutions \textit{agree} up to order $O(r^{-4})$. This will be called the EH branch or the $k=1$ case, and
      \item Eqs.(\ref{LLk27d}) and (\ref{LLk27d2}) have solutions that share a branch, of second order, satisfying Eq.(\ref{AsymptAdSBehavior}) and have the same asymptotic behavior up to order $O(r^{-1})$. This corresponds to the $k=2$ solution discussed in \cite{Crisostomo:2000bb} for constant transversal curvature.
\end{itemize}

In what follows the analysis of the nonconstant curvature transverse sections will be discussed. Specifically, some new solutions will be classified according to their degeneration on their effective cosmological constants. Afterward, the constraints on the $c(i)$'s necessary to ensure sound action principles for each of the solutions will be studied.

\subsection{Einstein like solutions}\label{EinsteinlikeSolutionSubs}

 Firstly, the solutions that share the asymptotic behavior of General Relativity in $d=7$ will be analyzed. These will be nicknamed Einstein-like Solutions and the corresponding EOM are \ref{LL7Dk1}) and (\ref{LLk17d2}) respectively. Later, it will be determined the constraints for these solutions to have sound action principles.

\subsubsection{\bf Gauss Bonnet in seven dimensions}
The solution to Gauss-Bonnet gravity, Eq.\ref{LL7Dk1}),  sharing one branch with EH can be extracted from the algebraic expression Eq.(\ref{EOMAdSSolved}), \textit{i.e.},
\begin{eqnarray*}
 K_{0} \left( c(0)\left(\frac{r^{2}}{l^{2}}-f(r)\right) + c(1)\right) \frac{r^{4}}{l^{4}} &+&  \\
 K_{1}\left(-c(0)\left(\frac{r^{2}}{l^{2}}-f(r)\right) f(r) + c(1)\left(\frac{r^{2}}{l^{2}}-2 f(r)\right) + c(2) \right) \frac{ r^{2}}{l^{2}} &=&   M.
\end{eqnarray*}
The solution has already been discussed in section \ref{GaussBonnetSection} and therefore there is not much ado but to recall that this solution only has finite conserved charges and action principle provided $c(1)^2 = c(2)c(0)$ is satisfied. As mentioned in section \ref{GaussBonnetSection} this constraint is also mandatory for higher dimensions solutions of GB gravity as well.

\subsubsection{\bf Cubic gravity in seven dimensions}
In $d=7$  one can consider as well Eqs. (\ref{LLk17d2}), which corresponds to $q=3$, a cubic gravity that shares one branch with EH. As before, the solution can be extracted from Eq.(\ref{EOMAdSSolved}), yielding the cubic equation for $f(r)$,
\begin{eqnarray*}
 K_{0} \left( c(0)\left(\frac{r^{2}}{l^{2}}-f(r)\right) + c(1)\right) \frac{r^{4}}{l^{4}} &+&   \\ K_{1}\left(-c(0)\left(\frac{r^{2}}{l^{2}}-f(r)\right) f(r)  + c(1)\left(\frac{r^{2}}{l^{2}}-2 f(r)\right) + c(2) \right) \frac{ r^{2}}{l^{2}}  &+& \\
K_{2}\left(c(0)\left(\frac{r^{2}}{l^{2}}-f(r)\right) f(r)^{2}-2 c(1) \left(\frac{r^{2}}{l^{2}}-\frac{1}{2}f(r)\right) f(r) +c(2)\left(\frac{r^{2}}{l^{2}}-3f(r)\right) + c(3)\right) &=& M.
\end{eqnarray*}
It is noteworthy to mention that the integration constant has been split into $M$ and $K_2c(3)$. In $d=7$ this is merely an artifact that maintains the general form of $d>7$ where $K_2c(3)$ has real meaning.

In this case, even though the exact expression of $f(r)$ can be obtained explicitly that will be omitted because that is cumbersome and shed no light on the discussion. Fortunately, the asymptotic form of $f(r)$ contains enough information to address the problem of finiteness. That asymptotic form is given by
\begin{equation}\label{d7k1q3}
  \lim_{r\rightarrow \infty} \sim \frac{c(1)}{c(0)} + \frac{r^2}{l^2} + (c(1)^2 -c(2)c(0))\frac{A}{r^2} +\frac{B}{r^{4}} + \ldots
\end{equation}
Here $B$ is a constant depending on $K_i$'s. $A$ is proportional to $M$ and a function of $c(i)$'s and $K_i$'s.

One can observe that order $r^{-2}$ spoils the corresponding EH asymptotia, namely $r^{d-3} \sim r^{-4}$, unless $c(1)^2 = c(2)c(0)$. This constraint is reinforced by checking that the conserved charges
\begin{eqnarray*}
    \lim_{r \rightarrow \infty} Q(\partial_t) &\approx& (c\! \left(1\right)^{2}-c\! \left(0\right) c\! \left(2\right))\left( -c\! \left(0\right) \left(K\! \left(0\right)-K\! \left(1\right)+K\! \left(2\right)\right) A + \frac{l^{2} \left(3 K\! \left(0\right)-7 K\! \left(1\right)+11 K\! \left(2\right)\right)}{c(0)^2} \right)  r^3  \\
    &-&  Mc\! \left(0\right) \left(K\! \left(0\right)-K\! \left(1\right)+K\! \left(2\right)\right) \\
    &+& \frac{c\! \left(1\right) l^{4} \left(-3 K\! \left(0\right)+7 K\! \left(1\right)-35 K\! \left(2\right)\right) \left(3 c\! \left(0\right) c\! \left(2\right)-2 c\! \left(1\right)^{2}\right)}{24 c\! \left(0\right)^{2}}
\end{eqnarray*}
and the action principle,
\begin{eqnarray*}
I &=& \lim_{r\rightarrow \infty}\beta  \left(c\! \left(1\right)^{2}-c\! \left(0\right) c\! \left(2\right)\right) \left(\frac{\left(K\! \left(0\right)-K\! \left(1\right)+K\! \left(2\right)\right) l^{2}}{4 c\! \left(0\right)}\right)r^2 \\
&+& -\frac{3 \beta l^{4} \left(\left(-\frac{2 K\left(0\right)}{3}+\frac{14 K\left(1\right)}{9}-\frac{22 K\left(2\right)}{9}\right) c\! \left(1\right)^{3}+c\! \left(0\right) c\! \left(2\right) \left(K\! \left(0\right)-\frac{7 K\left(1\right)}{3}+\frac{11 K\left(2\right)}{3}\right) c\! \left(1\right)+\frac{8 c\left(3\right) c\left(0\right)^{2} K\left(2\right)}{3}\right)}{8 c\! \left(0\right)^{2}} \\
&-& P(r_+),
\end{eqnarray*}
are both finite provided $c(1)^2 = c(2)c(0)$.

To finish this discussion it must be emphasized that in this case there is no restriction on $c(3)$, as occurred for $c(2)$ in $d=5$.

\subsubsection{ \bf Higher Dimensions}

The higher dimensional ($d>7$) extension of the solution above can be done directly from Eq.(\ref{EOMAdSSolved}). Its asymptotic form, for $d>7$, is given by
\begin{equation}\label{d>7k1q3}
  \lim_{r\rightarrow \infty} f(r) \sim \frac{c(1)}{c(0)} + \frac{r^2}{l^2} + (c(1)^2 -c(2)c(0))\frac{A}{r^2} + \left((c(3) -\frac{c(1)^3}{c(0)^2}\right)\frac{B}{r^4} +\frac{D}{r^{d-3}} + \ldots
\end{equation}
It can be observed that this expression differs from the expected EH behavior $(r^{-(d-3)})$ by two terms. This first one $(c(1)^2 -c(2)c(0)=0$ is the same as in $d>5$. The second, for $d>7$, is $c(3)c(0)^2 = c(1)^3$. It is direct, but cumbersome, to check that the finiteness of the conserved charges and action principle requires $c(1)^2=c(2)c(0)$ and $c(3)c(0)^2 = c(1)^3$.

\subsubsection{ \bf Final Comments}

It is worth noticing that each of the conditions mentioned above is trivially satisfied if a constant curvature transverse section were considered. Furthermore, it is not clear that any other than a constant curvature manifold can satisfy them.

The generality of the constraints $c(1)^2 = c(2)c(0)$ and  $c(3)c(0)^2 = c(1)^3$, being valid for GR-GB and cubic gravity in $d=7$ may be foreseen, within the Einstein branch, the rise of further constraints as the higher order on $R$ in the Lovelock theory increases. In this way, quartic gravity would require a constraint of $c(4)$ for $d>9$, and so on. Unfortunately, in general, this cannot be confirmed analytically because the equation for $f(r)$ cannot be solved for powers higher than four.

\subsection{Second order degenerated solutions}
Solutions presenting a second-order degeneration on the ground state, \textit{i.e.}, the solutions of \eqref{LLk27d} will be called second-order degenerated solutions. Before proceeding it is worth recalling in the case of a constant curvature transverse section the slope is given by $O(r^{-1})$ \cite{Crisostomo:2000bb}.

\subsubsection{ \bf Seven Dimensions}
The static solution to Eqs. (\ref{LLk27d}) can be obtained, see Eq.(\ref{EOMAdSSolved}), by solving the algebraic relation
\begin{equation}\label{algebraicf72k2}
  K_0\left(c(0)\left(\frac{r^2}{l^2} -f(r)\right)^2 + 2 c(1)\left(\frac{r^2}{l^2} -f(r)\right) + c(2) \right) \frac{r^2}{l^2} = M.
\end{equation}
The physical solution is given by
\begin{equation}\label{f72k2}
  f(r) = \frac{c(1)}{c(0)} + \frac{r^2}{l^2} - \sqrt{\frac{M l^2}{{K_0c(0) r^2}} + \frac{1}{c(0)^2} (c(1)^2 - c(2)c(0))}
\end{equation}
This solution presents two clear cases,  $c(1)^2 = c(2)c(0)$ and  $c(1)^2 \neq c(2)c(0)$.

\begin{itemize}
    \item For  $c(1)^2 = c(2)c(0)$ the solution is given by
\[
f(r) =  \frac{c(1)}{c(0)} + \frac{r^2}{l^2} -\frac{1}{r} \sqrt{ \frac{Ml^2}{K_0 c(0)} }
\]
which shares the slope $r^{-1}$ of the constant curvature transverse section solution \cite{Crisostomo:2000bb,Aros:2000ij}. The conserved charge is finite and given by
\begin{equation}\label{7k2q2}
  Q(\partial_t) = Ml^4 + l^4 \frac{K_0 c(1)^3}{6 c(0)^2},
\end{equation}
connecting $M$, the integration constant, with the mass/energy of the solution. In the same fashion, it is direct to check that the action principle is also finite and given by
\[
I = \beta \left(l^4 \frac{K_0 c(1)^3}{6 c(0)^2}\right) - P(r_{+}),
\]
where $P(r_{+})$ is finite and $\beta$ is the inverse of the Euclidean period.
\item The case  $c(1)^2 \neq c(2)c(0)$ presents some unique features that can observed in
\begin{equation}
\lim_{r \rightarrow \infty} f(r) \sim   \frac{c(1)}{c(0)} - \sqrt{\frac{c(1)^2 - c(2)c(0)}{c(0)^2}} + \frac{r^2}{l^2} - \frac{M}{r^2} \frac{l^2}{2 K_0\sqrt{c(1)^2 - c(2)c(0)}} + M^2 \frac{C_4}{r^4} + \ldots.,
\end{equation}
where $C_4$ is a constant depending on $(c(1)^2 - c(2)c(0))^{-3/2}$. One can notice that the leading order $O(r^{-1})$ has been replaced by $O(r^{-2})$, however this does not affect the existence of a finite conserved charge $Q(\partial_t)$, which is given by
\begin{equation}\label{7k2q2Version2}
  Q(\partial_t) = M l^2 + Ev,
\end{equation}
with
\begin{eqnarray*}
  Ev &=& \frac{K_0 l^{4}}{3 c(0)^2 }\left(c(1)^2 - c(2)c(0) \right)^{\frac{3}{2}} \\
   &+& \frac{K_0 l^{4}}{6 c(0)^2} c(1) \left( 3 c(2)c(0) - 2 c(1)^2 \right)
   \end{eqnarray*}
establishing that $M$, the integration constant in Eq.(\ref{algebraicf72k2}), is related with the mass/energy of the solution. Furthermore, $Ev$ has a soft limit to the previous case. FThe action principle is finite as well and is given by
\begin{eqnarray}
  I &=& \beta \left(\frac{K_0 l^{4}}{3 c(0)^2 }\left(c(1)^2 - c(2)c(0) \right)^{\frac{3}{2}} \right. \label{Action7kk2q2versionI} \\
   &+& \left. \frac{K_0 l^{4}}{6 c(0)^2} c(1) \left( 3 c(2)c(0) - 2 c(1)^2 \right)\right) - P(r_+) \nonumber
\end{eqnarray}

\end{itemize}

\subsubsection{\textbf{Higher dimensions}}

In dimensions $d>7$, one can observe the same basic features observed in 7 dimensions, with the sole exception that the coefficients $c(i)$, with $i>2$, play a role in the renormalization processes. The general solution is given by
\[
f(r) = \frac{c(1)}{c(0)} + \frac{r^2}{l^2} - \sqrt{\frac{M l^{d-5}}{{K_0c(0) r^{d-5}}} + \frac{1}{c(0)^2} (c(1)^2 - c(2)c(0))}
\]
As before, the asymptotic behavior splits according  $c(1)^2 = c(2)c(0)$ or not.
\begin{itemize}
\item For $c(1)^2 = c(2)c(0)$ the solution is given by
\[
f(r) = \frac{c(1)}{c(0)} + \frac{r^2}{l^2} - \sqrt{\frac{M l^{d-3}}{{K_0c(0) r^{d-3}}}},
\]
which matches the asymptotia of the constant curvature transverse section case. For this solution, though it is a bit long, it can be demonstrated that it has a finite conserved charge
\[
Q(\partial_t) = Ml^2 +  \frac{l^{d-3} K_{0} c\left(1\right)^{(d-1)/2}}{ 2^{(d-1)/2}c\! \left(0\right)^{(d-3)/2}}
\]
The action principle is also finite and given by
\[
I = \beta \frac{l^{d-3} K_{0} c(1)^{(d-1)/2}}{ 2^{(d-1)/2}c\! \left(0\right)^{(d-3)/2}} - P(r_+).
\]
where $\beta$ is the Euclidean period. These results only confirm, as expected, that $c(1)^2 = c(2)c(0)$ yields an analogous situation as the corresponding constant curvature constant transverse section solution in \cite{Aros:2000ij}.

\item The case $c(1)^2 \neq c(2)c(0)$ has the different asymptotic behavior given by
\[
\lim_{r \rightarrow \infty} f(r) =  \frac{c(1)}{c(0)} - \sqrt{\frac{c(1)^2 - c(2)c(0)}{c(0)^2}} + \frac{r^2}{l^2} - \frac{M}{2 K_0\sqrt{c(1)^2 - c(2)c(0)}}\left(\frac{l^{d-5}}{r^{d-5}}\right) + M^2 C_{d-2} \left( \frac{ l^{d-3}}{r^{d-3}}\right) + \ldots.
\]
Regardless of this completely different behavior, still the associated conserved charge is finite providing some relations between the $c(i)$, $i>2$, are satisfied. Remarkably, these relations cannot be satisfied by a constant curvature transverse section, and therefore these solutions represent a completely new family of well-defined solutions. If those restrictions are satisfied then the conserved charge is given by
\[
Q(\partial_t) =  Ml^{d-3} + E_{v}
\]
where $E_{v}$ is a finite, but cumbersome, function of $c(1)$ and $c(2)$. For instance, in $d=11$ the finiteness requires that
\begin{eqnarray*}
    c(3) &=& -\frac{2 c\! \left(0\right)^{2} \sqrt{\frac{c\left(1\right)^{2}-c\left(2\right) c\left(0\right)}{c\left(0\right)^{2}}}\, c\! \left(2\right)-2 c\! \left(1\right)^{2} c\! \left(0\right) \sqrt{\frac{c\left(1\right)^{2}-c\left(2\right) c\left(0\right)}{c\left(0\right)^{2}}}-3 c\! \left(1\right) c\! \left(0\right) c\! \left(2\right)+2 c\! \left(1\right)^{3}}{c\! \left(0\right)^{2}}\\
    c(4) &=& \frac{-8 \sqrt{\frac{c\left(1\right)^{2}-c\left(2\right) c\left(0\right)}{c\left(0\right)^{2}}}\, c\! \left(2\right) c\! \left(1\right) -3 c\! \left(2\right)^{2}}{c\! \left(0\right)}+\frac{4 c\! \left(1\right)^{2} \left(2 c\! \left(1\right) \sqrt{\frac{c\left(1\right)^{2}-c\left(2\right) c\left(0\right)}{c\left(0\right)^{2}}}+3 c\! \left(2\right)\right)}{c\! \left(0\right)^{2}}-\frac{8 c\! \left(1\right)^{4}}{c\! \left(0\right)^{3}}
\end{eqnarray*}
which yields
\begin{eqnarray*}
       E_{v} &=& \left(\left(-\frac{15 c\! \left(1\right)}{32 c\! \left(0\right)^{2}}+\frac{\sqrt{c\! \left(1\right)^{2}-c\! \left(2\right) c\! \left(0\right)}}{8 c\! \left(0\right)^{2}}\right) c\! \left(2\right)^{2} \right.\\
       &+& \left(\frac{5 c\! \left(1\right)^{3}}{4 c\! \left(0\right)^{3}}-\frac{7 \sqrt{c\! \left(1\right)^{2}-c\! \left(2\right) c\! \left(0\right)}\, c\! \left(1\right)^{2}}{8 c\! \left(0\right)^{3}}\right) c\! \left(2\right) \\
       &-&\left.\frac{3 c\! \left(1\right)^{5}}{4 c\! \left(0\right)^{4}}+\frac{3 \sqrt{c\! \left(1\right)^{2}-c\! \left(2\right) c\! \left(0\right)}\, c\! \left(1\right)^{4}}{4 c\! \left(0\right)^{4}}  \right) K\! \left(0\right) l^{8}
\end{eqnarray*}
By the same token, in $d=11$ the action principle is also finite and given by
\begin{eqnarray*}
       I &=& \beta\left(\left(-\frac{15 c\! \left(1\right)}{32 c\! \left(0\right)^{2}}+\frac{\sqrt{c\! \left(1\right)^{2}-c\! \left(2\right) c\! \left(0\right)}}{8 c\! \left(0\right)^{2}}\right) c\! \left(2\right)^{2} \right.\\
       &+& \left(\frac{5 c\! \left(1\right)^{3}}{4 c\! \left(0\right)^{3}}-\frac{7 \sqrt{c\! \left(1\right)^{2}-c\! \left(2\right) c\! \left(0\right)}\, c\! \left(1\right)^{2}}{8 c\! \left(0\right)^{3}}\right) c\! \left(2\right) \\
       &-&\left.\frac{3 c\! \left(1\right)^{5}}{4 c\! \left(0\right)^{4}}+\frac{3 \sqrt{c\! \left(1\right)^{2}-c\! \left(2\right) c\! \left(0\right)}\, c\! \left(1\right)^{4}}{4 c\! \left(0\right)^{4}}  \right) K\! \left(0\right) l^{8} \\
       &-& P(r_+)
\end{eqnarray*}

\end{itemize}

\subsection{Second Order like Solutions in seven dimensions}

It will be understood by \textit{second order-like degenerated solutions} those with the same asymptotic behavior of second-order degeneration, namely the previous case, but whose equations of motion have $R^3$, or higher, powers on the Riemann tensor ( {\it i.e} at least cubic gravity) and have nonconstant transversal curvature. {This case is described by the equation \eqref{LLk27d2}.} This can occur only for $d\geq 7$. On the other hand, unfortunately, in general, the exact expression of $f(r)$ cannot be obtained for $d>8$, leaving only $d=7,8$ as sound options. In this section only cubic gravity with $k=2$ in $d=7$ will be explored as $d=8$ with $k=2$ essentially has the features. In $d=7$ the solution  satisfies the algebraic equation, see Eq.(\ref{EOMAdSSolved}),
\begin{eqnarray}
    K_0\left(c(0)\left(\frac{r^2}{l^2} -f(r)\right)^2 + 2 c(1)\left(\frac{r^2}{l^2} -f(r)\right) + c(2) \right) \frac{r^2}{l^2} &+& \label{K23Eq} \\
  K_1 \left(-c(0) \left(\frac{r^{2}}{l^{2}}-f(r)\right)^{2} f(r)-2 c(1) \left(\frac{r^{2}}{l^{2}}-f(r)\right) f(r) +c(1) \left(\frac{r^{2}}{l^{2}}-f(r)\right)^{2} +c(2) \left(2\frac{r^{2}}{l^{2}}-3 f(r)\right)+c(3)\right) &=& Z_0 \nonumber.
\end{eqnarray}
As in some previous cases, here the integration constants have been split into $K_1c(3)$ and $Z_0$ just to preserve the form of the solution for $d>7$. It must be emphasized that the solution exists for $d>8$ provided the higher power of the Riemann tensor is not increased.

Unfortunately, the exact form of $f(r)$, in this case, is remarkably cumbersome and thus does not provide any hindsight. However, its asymptotic form provides enough information to perform the analysis. As previously, there is a split between the cases  $c(1)^2 = c(2)c(0)$ and  $c(1)^2 \neq c(2)c(0)$.
\begin{itemize}
  \item For $c(1)^2 = c(2)c(0)$ the asymptotic form is given by
\begin{equation}\label{fasb73k2go}
  \lim_{r \rightarrow \infty} f(r) \sim \frac{c(1)}{c(0)} + \frac{r^2}{l^2} - \left(\frac{l}{r}\right)\left(  \sqrt{ \frac{M}{c\! \left(0\right) \left(K\left(0\right)-K\left(1\right)\right)}} \right) + \ldots
\end{equation}
It is direct to check that this asymptotic form yields a finite action principle and finite conserved charge. One can notice that the presence of $c(3)c(0)^2-c(1)^3$, previously noticed in subsection \ref{EinsteinlikeSolutionSubs} as a constraint to ensure an Einstein-like behavior, in this case however it only shifts the value of the effective mass as
\begin{equation}\label{Massd7K2q3}
  Q(\partial_t) = M \,l^{4}+\frac{\left(K\! \left(0\right)-7 K\! \left(1\right)\right) l^{4} c\! \left(1\right)^{3}}{6 c\! \left(0\right)^{2}}
\end{equation}
The action principle is given by
\begin{equation}\label{ActionPrincipled7K2q3}
  I = \beta \left(\frac{\left(c(1)^{3} (K(0)-K(1))-6 c\! \left(3\right) c\! \left(0\right)^{2} K\! \left(1\right)\right) l^{4}}{6 c\! \left(0\right)^{2}}\right)- P(r_+)
\end{equation}
It is direct to notice that this reproduces the features of the constant curvature transverse section solution.

\item For  $c(1)^2 \neq c(2)c(0)$  the asymptotic form is given by
  \begin{equation}\label{fasb73k2notgo}
  \lim_{r \rightarrow \infty} f(r) \sim \frac{c(1)}{c(0)} - \sqrt{\frac{c(1)^2 - c(2)c(0)}{c(0)^2}}  + \frac{r^2}{l^2} + \left(\frac{l}{r}\right)^2 C_2(M,c(i),K_i) + \left(\frac{l}{r}\right)^4 C_4(M,c(i),K_i) \ldots
\end{equation}
where
\begin{equation}
    C_2(Z_0,c(i),K_i) =  \frac{1}{2\kappa^2 \sqrt{c(0)}}\left( \Upsilon + \frac{Z_0}{2\Xi^3}\right)^2
\end{equation}
with
\begin{eqnarray*}
    \kappa &=& K_1 - K_0,\\
    \Xi^4 &=& c(2) - \frac{c(1)^2}{c(0)}, \text{ and  }  \\
    \Upsilon &=&  \frac{\Xi^2}{\sqrt{c(0)}} - \frac{3}{2} \left( \Xi - \frac{c(1)}{c(0) \Xi^3}\right).
\end{eqnarray*}
In Eq.(\ref{fasb73k2notgo}), $C_4(Z_0,c(i),K_i)$ is also a function depending on the constants $Z_0,c(0),c(1), c(2), K_0$ and $K_1$. In this case, the conserved charge is given
\begin{equation}\label{Massd7K2q3V2}
Q(\partial_t) =  M l^4 + E_v
\end{equation}
with
\[
M =  \frac{\left(2\Upsilon \Xi^{3} K(1)+ Z_0 \right)^{2}}{4 \kappa  \,\Xi^{4}}.
\]
One can notice, see Eq.(\ref{K23Eq}), that defining $M$ can be done in general and adjusted by fine tuning $Z_0$. Here,
\[
E_v = \left(-\frac{c\! \left(2\right) l^{4}}{3 \sqrt{c\! \left(0\right)}}+\frac{c\! \left(1\right)^{2} l^{4}}{3 c\! \left(0\right)^{\frac{3}{2}}}\right) \Xi^{2}+\frac{c\! \left(1\right) c\! \left(2\right) l^{4}}{2 c\! \left(0\right)}-\frac{c\! \left(1\right)^{3} l^{4}}{3 c\! \left(0\right)^{2}}
\]

\begin{equation}\label{ActionPrincipled7K2q3V2}
  I = \beta \left( \frac{l^{6} \left(c\! \left(2\right) c\! \left(0\right)-c\! \left(1\right)^{2}\right)^{2} K\! \left(0\right)^{2}}{3 \sqrt{l^{4} K\! \left(0\right)^{2} \left(c\! \left(1\right)^{2}-c\! \left(2\right) c\! \left(0\right)\right)}\, c\! \left(0\right)^{2}}+\frac{\left(\frac{3 c\left(0\right) c\left(1\right) c\left(2\right)}{2}-c\! \left(1\right)^{3}\right) l^{4} K\! \left(0\right)}{3 c\! \left(0\right)^{2}} \right) - P(r_+).
\end{equation}

\end{itemize}

\subsection{Chern Simons in seven dimensions}

Besides the two cases above, in $d=7$ a cubic Lagrangian could correspond to Chern-Simons gravity. In this case, the EOM is given by Eq.(\ref{CS7d}) which reduces the static ansatz to
\begin{equation}\label{CSsolution7d}
   \left(\frac{r^{2}}{l^{2}}-f\right)^{3} c(0)+6 \left(\frac{r^{2}}{l^{2}}-f\right)^{2} c\left(1\right)+6 \left(\frac{r^{2}}{l^{2}}-f\right)c(2) = C \cdot c(0),
\end{equation}
where $C\,c(0)$ is an integration constant related with the \textit{energy} of the solution. This yields
\begin{equation}\label{CS7dSol}
  f(r) = \frac{c(1)}{c(0)} + \frac{r^2}{l^2} - C_2.
\end{equation}
The exact form of $C_2 = C_2(c(q),C)$ is not illustrative, but can be reckoned from solving
\[
  - \left(\frac{c(1)}{c(0)} - C_2\right)^{3} c(0) + 6 \left(\frac{c(1)}{c(0)}- C_2 \right)^{2} c(1)- 6 \left(\frac{c(1)}{c(0)}- C_2 \right)c(2)=C \, c(0).
\]
Eq.(\ref{CS7dSol}) essentially can be cast as a generalization of the constant curvature case discussed in \cite{Crisostomo:2000bb}. It can be inferred that for Chern Simons gravity in $d=7$ there are no constraints so that the solution with nonconstant transversal curvature has finite value for its conserved charge due to its energy being related to $C_2$ and consequently for its action.

\section{Discussion}

In this work it has been shown some new proper asymptotically AdS static black hole solutions with nonconstant curvature transverse sections. The analysis was carried out by establishing the conditions that yield the finiteness of the conserved charges, associated with the time symmetry, and the corresponding action principle. These solutions satisfy Lovelock gravities EOM given by Eq.(\ref{EOMAdSForm}), whose form can be sketched as
\[
\left(R + \frac{e^2}{l^2}\right)^k \left(\sum_{s=0}^{q-k} K(s) R^s \left(\frac{e}{l}\right)^{2(q-k-s)}\right)=0,
\]
with $q < [(d-1)/2]$. Here $q$ represents the highest power of the Riemann tensors in the Lagrangian, and $d$ the dimension. Part of the analysis revealed that the solutions can be classified according to their degeneration around the AdS ground state, given by $k$ in the equation above. In particular, solutions with the degenerations up third order are discussed in detail for $d \leq 7$. The results can be summarized as follows
\begin{itemize}
  \item For Chern Simons gravity, namely $d=2n+1$ and $(k=q=n)$, the situation is essentially identically to the constant curvature transverse section case with no constraint on the $c(q)'s$. The effect considering non-constant curvature transverse sections is a modification in the mass/energy by a function of the corresponding $c(0),c(1),\ldots c(n-1)$.
  \item For $(k,q)=(1,1)$, namely the solutions of General Relativity, it is only required to know $c(0)$ and $c(1)$, and thus no further constraints on the values of $c(q)$, for $q>2$ arises. This grants GR a singular status where transverse sections are essentially unconstrained. This is a very appealing condition in the context of the AdS/CFT conjecture, as mentioned above.
  \item For $k=1$ and $ 1 < q < [d/2]$ (called Einstein-like solutions), meaning theories with higher power of the Riemann tensor that have branches with a solution that could share the asymptotia of General Relativity, the finiteness of the conserved charges and the action principle impose an increasing number conditions, with $q$, on the $c(p)'s$. For instance, $q=2$ it is required that
\[
c(1)^2 = c(2)c(0)
\]
in $d > 5$. In the same fashion, for $d \geq 9$ and $q \geq 3$ it is also required that
\[
c(3)c(0)^2=c(1)^3.
\]
These two conditions are satisfied by any constant curvature transverse sections. For higher orders in $q$ and higher dimensions, one can foresee the rise of a set of conditions of the same fashion to be trivially satisfied by any constant curvature transverse sections.

\item For $(k=2,q=2)$, {called Second order degenerated solutions,} the solutions split into the two disjointed cases $c(1)^2 = c(2)c(0)$ and $c(1)^2 \neq c(2)c(0)$. In the first case, the slope agrees with the one given by considering the constant curvature transverse section case and no further constraints arise.

Remarkably, for  $c(1)^2 \neq c(2)c(0)$ both the charges and action principle can be finite, but the slope of the solutions differ from the constant curvature transverse section case. This introduces a new family of solutions whose extension, for instance to stationary solutions, can yield some interesting new results. This will be explored elsewhere.

\item  For $k=2$ and $ 2 < q < [d/2]$, (called second-order like solutions) the behavior of $(k,q)=(2,2)$ is recovered completely. For instance, the split between the behaviors of $c(1)^2 c(2)c(0)$ and $c(1)^2 \neq c(2)c(0)$ cases is reproduced.

\end{itemize}

To finish it is worth mentioning just a few interesting open questions. The analysis of the thermodynamics requires addressing if it is possible to have vanishing temperature solutions and if some constrains on the $c_q$ coefficients rise. Unfortunately, this requires determining the specific form of $f(r)$ which, as mentioned before, is not possible in general for $d>8$. This is not an easy task as even the constant curvature transverse section solutions of GB gravity require a thorough discussion \cite{Bravo-Gaete:2021hza}. Finally, as mentioned before, in this work only a particular method of regularization was used, and thus it is also an open question if another method could yield different constraints on the $c(q)$ or any at all.

This work has discussed the physical viability of spaces with non-constant curvature, prompting consideration of potential future applications. In this regard, it is known, for example, that by employing the Kerr-Schild ansatz, rotating solutions can be obtained. This approach involves applying a perturbation to a seed spacetime. Thus far, such seed spaces have primarily featured constant curvature in the literature. For instance, in references \cite{Anabalon:2009kq,Fierro:2020wps}, rotating solutions in EGB gravity were obtained by perturbing an AdS (and also flat \cite{Anabalon:2009kq}) spacetime in ellipsoidal coordinates. See also references \cite{Adair:2020vso,Cvetic:2016sow}. This raises the question of whether in a future study, a mechanism analogous to Kerr-Schild could be designed where the seed possesses non-constant curvature. However, the latter seems to be a very difficult task, and a comprehensive analysis is necessary to propose an ansatz for stationary geometries. The Kerr-Schild ansatz big advantage is to allow visualizing Killing vectors, both light-like and space-like manifestly, and thus to implement the axial symmetry beforehand. See for instance \cite{Malek:2014dta} for a review. Unfortunately, the expression of the fundamental equations in this work, eq (\ref{LEOM}), in terms of a Kerr-Shield ansatz does not seem to be trivial, or generic, seeming to need to be studied case by case ($n,k$). Even though this is a promising line of study, it is certainly beyond this work's scope.

\acknowledgments

This work was partially funded through FONDECYT-Chile 1220335. Milko Estrada is funded by the FONDECYT Iniciaci\'on Grant 11230247.

\appendix

\section{Lovelock equations of motion}\label{LovelockEOM}

The variation of Eq,(\ref{initialaction}) is given by,
\begin{equation}\label{Variation}
\delta_{0}{\mathbf{L}} = G_f\delta_{0} e^f + D(\delta_{0} w^{ab}) \tau_{ab}
\end{equation}
where
\begin{equation}
\tau_{ab}=\frac{\partial{\mathbf{L}}}{\partial R^{ab}} = \kappa\sum_{p=0}^{n-1} p \alpha_p
\,\epsilon_{ab a_3\ldots a_{2n}} \left[(R)^{p-1} \left(\frac{e}{l}\right)^{2n-2p}\right]^{a_3
\ldots a_{2n}}\label{firstvariation}.
\end{equation}
and
\begin{equation}\label{GraviationalEOM}
G_f=\frac{\overleftarrow{\partial}{\mathbf{L}}}{\partial e^f} = \kappa\sum_{p=0}^{n-1}
(2n-2p)\alpha_p \,\epsilon_{a_1\ldots a_{2n-1}f} \left[(R)^{p}
\left(\frac{e}{l}\right)^{2n-2p-1}\right]^{a_1 \ldots a_{2n-1}}
\end{equation}
Using the Stoke theorem the second term in Eq.(\ref{Variation}) can be split such that obtaining the second field equation $D(\tau_{ab})=0 $ and the boundary term
\begin{equation}\label{GraviationalTheta}
\Theta(\delta_{0} w^{ab} w^{ab} e^d)=\kappa\sum_{p=0}^{n-1} p \alpha_p \,\epsilon_{a_1\ldots
a_{2n}} \left[ \delta_{0} w (R)^{p-1} \left(\frac{e}{l}\right)^{2n-2p}\right]^{a_1 \ldots a_{2n}}.
\end{equation}

It is worth mentioning that for general $\alpha_p$,  $T^{a}=de^{a}+ w^{a}_{\,\,\,b}e^{b}=0$  is the only solution for $D(\tau_{ab})=0$, thus this formalism is equivalent to the metric formalism. However, for the special case of Chern-Simons, although $T^a=0$ is a solution, it is not the most general one. See for instance \cite{Troncoso:1999pk}.

\section{Regularization}\label{OddDReg}

The renormalization process of Lovelock gravity in even dimensions ($d \ge 4$) is direct and can be accomplished by adding the corresponding Euler density with a suitable unitless constant. See for instance \cite{Aros:1999kt,Kofinas:2008ub}. Because of this, and for simplicity, only the renormalization process in odd dimensions will be sketched. For further details, see, \cite{Mora:2004kb,Mora:2004rx,Kofinas:2007ns,Miskovic:2007mg}. Unlike even dimensions, in this case, the regulation process must be carried by a suitable boundary term at the asymptotic AdS region. For the horizon, as no divergencies can arise, no additional term is necessary to attain finiteness.

The variation of the Lovelock action on the shell can be written as:
\begin{equation}\label{VariationLL}
    \delta_0 I_{LL} = \int_{\partial \mathcal{M}} l^{2n-1} \left( \sum_{p=0}^n p (-1)^{2n-2p+1} \alpha_p \right) \delta_0 \omega R^{n-1}.
\end{equation}
From this, it is straightforward to realize, as mentioned above, that there is not a proper set of boundary conditions that define $\delta I_{LL} =0$ as $R$ diverges in the asymptotically AdS region. This can be amended by the addition of the boundary term given by \cite{Mora:2004kb, Mora:2000ts}:
\begin{equation}\label{BoundaryOdd}
 I_{R}= \int_{\partial \mathcal{M}_\infty} B_{2n} = \kappa \int_{\partial \mathcal{M}_{\infty}} \int_{0}^1 \int_0^t \left(K e \left(\tilde{R} + t^2 (K)^2 + s^2 \frac{e^2}{l^2}\right)^{n-1} \right) ds dt
\end{equation}
where $\tilde{R}$ and $K$ stand for the Riemann two-form and extrinsic curvature one-form respectively of the boundary $\partial \mathcal{M}_\infty = \mathbb{R} \times \partial \Sigma_\infty $. One must recall the Gauss Codazzi decomposition:
\begin{equation}
    \left. \tilde{R}^{ab} + ((K)^2)^{ab} \right|_{\partial \mathcal{M}_{\infty}}=  \left. R^{ab} \right|_{\partial \mathcal{M}_{\infty}}
\end{equation}
where $R^{ab}$ is the Riemann two form of $\mathcal{M}$. $\kappa$ in Equation (\ref{BoundaryOdd}) stands for a constant to be determined. The variation of Equation (\ref{BoundaryOdd}) yields:
\begin{eqnarray}
\delta_0 I_{R} &=& \kappa  \int_{\partial \mathcal{M}_{\infty}} \int_{0}^1 \left(e \delta_0 K  - \delta e K_0\right) \left(\tilde{R} + t^2 (K)^2 + t^2 \frac{e^2}{l^2}\right)^{n-1} dt \\
 &+& \kappa n \int_{\partial \mathcal{M}_{\infty}} \int_{0}^1  \left( e \delta_0 K \left(\tilde{R} + (K)^2 + t^2 \frac{e^2}{l^2}\right)^{n-1} \right) dt \nonumber\\
\end{eqnarray}
For an asymptotically local AdS space, as the boundary is approached, it is satisfied that $e \delta_0 K  - \delta_0 e K \rightarrow 0$ and $\delta K \rightarrow \left. \delta \omega \right|_{\partial \mathcal{M}}$. The fundamental key for the computation, however, is the fact that $e^2 \rightarrow -l^2 R$. Finally, these conditions allow us to express variation as:
\begin{equation}
    \delta_0 I_{R} =   \kappa n \int_{\partial \mathcal{M}_{\infty}}    e \delta_0 K  R^{n-1} \left( \int_{0}^1 \left(1 - t^2 \right)^{n-1} \right) dt.
\end{equation}
In this way, the variation of $I = I_{LL} + I_{R}$:
\begin{equation}
\delta_0 I  = \int_{\partial \mathcal{M}_{\infty}} \left(\delta_0 K \left(\frac{e}{l}\right) R^{n-1} \right)  \left( l^{2n-1}\sum_{p=0}^n p (-1)^{2n-2p+1} \alpha_p  + n l \kappa \frac{\Gamma(n) \sqrt{\pi}}{2 \Gamma\left(n + \frac{1}{2}\right)}\right) + \ldots
\end{equation}
here, $\ldots$ stands for the integral of Equation (\ref{VariationLL}) on the horizon.  This defines:
\begin{equation}\label{KLovelockOdd}
    \kappa =  \frac{2 l^{2n-2}}{n}  \left(\sum_{p=0}^n p (-1)^{2n-2p} \alpha_p  \right) \frac{\Gamma\left(n + \frac{1}{2}\right)}{\Gamma(n) \sqrt{\pi}}
\end{equation}

In doing this now, there is a proper action principle. The Noether charge, in this case, is given by:
\begin{eqnarray}
  Q(\xi)_{\infty} &=& \int_{\partial \Sigma_{\infty}} \left( I_{\xi} \omega \left(\sum_{p=0}^n p \alpha_p R^{p-1} e^{2(n-p)+1} \right) \right. \label{NoetherOddFormal} \\
  &+& \left. \kappa I_{\xi} \left(\int_{0}^1 \int_0^t K e \left(\tilde{R} + t^2 (K)^2 + s^2 \frac{e^2}{l^2}\right)^{n-1} \right) ds dt \right)\nonumber
\end{eqnarray}
The direct evaluation of this expression for $\xi = \partial_t$ on the static spaces considered yields the final result.

To conclude this section, it is convenient to express the presymplectic form in terms of the regularized Noether charge and the variation of the action defined by the boundary term in Equation (\ref{BoundaryOdd}). This yields:
\begin{eqnarray}
  \left.\hat{\delta} G(\xi)\right|_{\infty} &=& \int_{\partial \Sigma_{\infty}}\hat{\delta} Q(\xi)_\infty + I_{\xi} \left( \kappa \int_{0}^{1} (e \hat{\delta}K - \hat{\delta} e K) \left(\tilde{R} + t^2 (K)^2 + t^2 \frac{e^2}{l^2}\right)^{n-1}  dt \right. \nonumber \\ &&+ 2\kappa(n-1) \hat{\delta}l \int_{0}^{1} \int_{0}^{1} K \left( \frac{e}{l} \right) \left(\tilde{R} + t^2 (K)^2 + s^2 \frac{e^2}{l^2}\right)^{n-1} ds dt \label{PresymplecticInOdd}\\ &&- 2\kappa(n-1) \hat{\delta}l \left. \int_{0}^{1} \int_{0}^{1} K \left( \frac{e}{l} \right)^{3} \left(\tilde{R} + t^2 (K)^2 + s^2 \frac{e^2}{l^2}\right)^{n-3} ds dt\right).\nonumber
\end{eqnarray}


\end{document}